\pdfoutput=1
\documentclass[11pt]{article}

\usepackage[preprint]{acl}
\usepackage{float}
\usepackage{multirow}
\usepackage{xspace}

\usepackage{pifont}%
\newcommand{\cmark}{\ding{51}}%
\newcommand{\xmark}{\ding{55}}%

\usepackage{times}
\usepackage{latexsym}
\usepackage{adjustbox}
\usepackage{amsmath}
\usepackage{booktabs}
\usepackage[inline]{enumitem}
\usepackage[most]{tcolorbox}
\usepackage{tcolorbox}
\usepackage{array}
\usepackage{graphicx} 
\usepackage{acronym}

\usepackage[T1]{fontenc}

\usepackage[utf8]{inputenc}

\usepackage{microtype}
\usepackage{pifont}
\usepackage{amssymb}
\usepackage{inconsolata}

\usepackage{graphicx}
\usepackage{algorithm}
\usepackage{algpseudocode}

\usepackage{amsmath, amssymb, algorithm, algpseudocode}

\newcommand{\OurData}{ClariMM\xspace}
\newcommand{\OurModel}{Mario\xspace}
\newcommand{\header}[1]{\vspace{2mm}\noindent\textbf{#1}}

\acrodef{CS}{conversational search}
\acrodef{LLM}{large language model}

\title{Multi-Turn Multi-Modal Question Clarification for Enhanced Conversational Understanding}

\author{
    \textbf{Kimia Ramezan}$^1$, \textbf{Alireza Amiri Bavandpour}$^1$, \textbf{Yifei Yuan}$^2$\thanks{Corresponding author} \\ 
    \textbf{Clemencia Siro}$^3$, \textbf{Mohammad Aliannejadi}$^3$ \\
    $^1$Sharif University of Technology \\
    $^2$University of Copenhagen \\
    $^3$University of Amsterdam \\
    \texttt{\{kim.ramezan03, alireza.amiri81\}@sharif.edu} \\
    \texttt{yiya@di.ku.dk} \\
    \texttt{\{c.n.siro, m.aliannejadi\}@uva.nl}
}

\begin{document}

\maketitle

\begin{abstract}

Conversational query clarification enables users to refine their search queries through interactive dialogue, improving search effectiveness. Traditional approaches rely on text-based clarifying questions, which often fail to capture complex user preferences, particularly those involving visual attributes. While recent work has explored single-turn multi-modal clarification with images alongside text, such methods do not fully support the progressive nature of user intent refinement over multiple turns. Motivated by this, we introduce the Multi-turn Multi-modal Clarifying Questions (MMCQ) task, which combines text and visual modalities to refine user queries in a multi-turn conversation. To facilitate this task, we create a large-scale dataset named \OurData{} comprising over 13k multi-turn interactions and 33k question-answer pairs containing multi-modal clarifying questions.
We propose \OurModel{}, a retrieval framework that employs a two-phase ranking strategy: initial retrieval with BM25, followed by a multi-modal generative re-ranking model that integrates textual and visual information from conversational history. Our experiments show that multi-turn multi-modal clarification outperforms uni-modal and single-turn approaches, improving MRR by 12.88\%. The gains are most significant in longer interactions, demonstrating the value of progressive refinement for complex queries.

\end{abstract}

\section{Introduction}

\Ac{CS} enables users and systems to collaboratively refine queries through dialogue~\cite{Radlinski2017ATF}, addressing limitations of traditional keyword-matching systems where single queries often fail to capture complete information needs~\cite{DBLP:journals/corr/abs-1907-06554,zamani-2020-generating}. Query clarification has emerged as a key mechanism for improving search accuracy by helping users refine ambiguous or incomplete queries~\cite{DBLP:journals/corr/abs-1901-05415,DBLP:journals/corr/abs-1805-04655}.

Current approaches to query clarification, while showing promise, face critical limitations in addressing complex information needs. Traditional systems rely predominantly on text-only clarifying questions~\cite{aliannejadidialogue,zamani-2020-generating}, proving insufficient when users need to understand or express preferences about visual characteristics. This limitation becomes inherent in certain query types or domains like healthcare (symptom identification), e-commerce (product selection), and technical support (problem diagnosis), where visual context is crucial for precise understanding~\citep{chiir-siro}.
\begin{figure}[t]
  \includegraphics[width=\columnwidth]{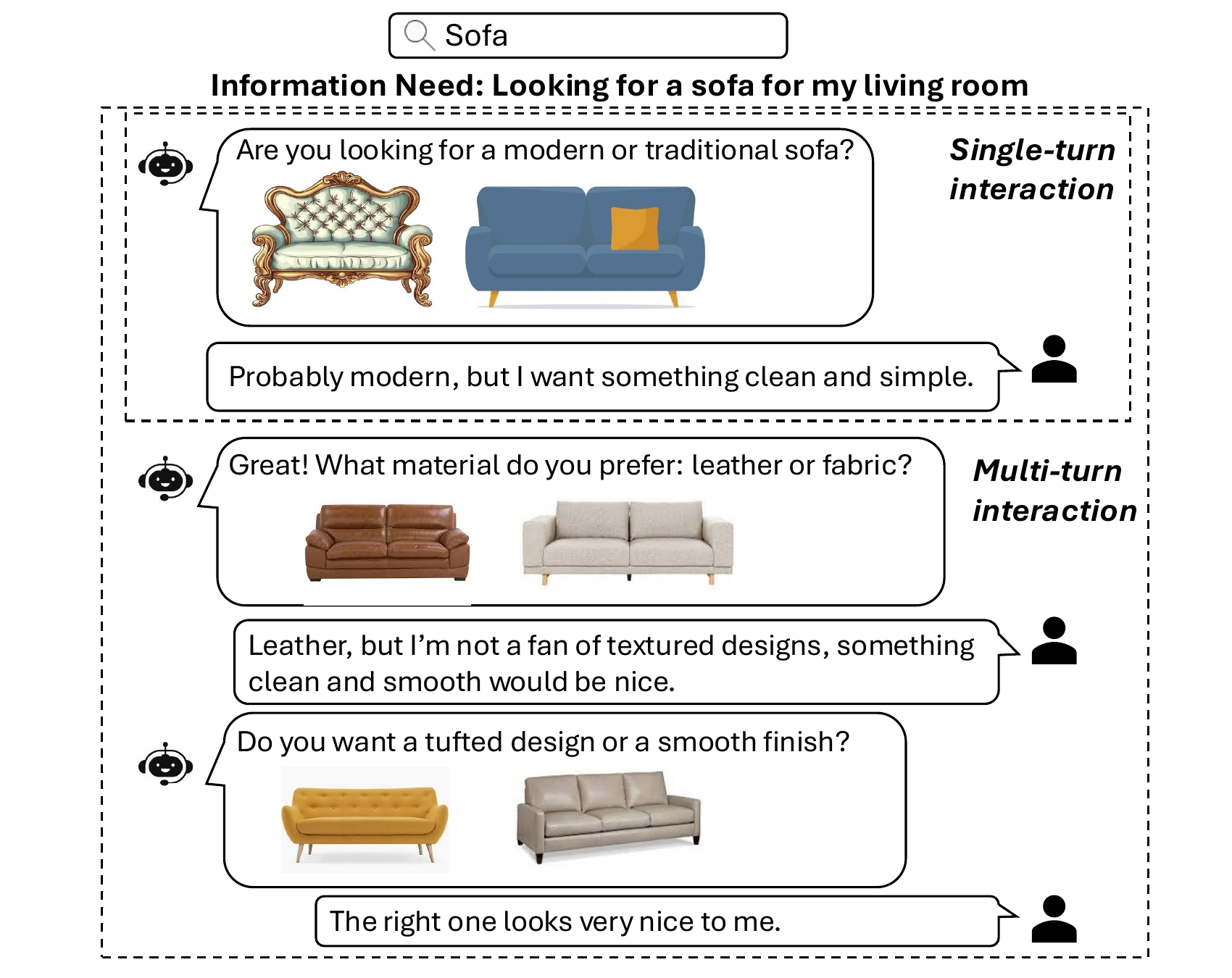}
  \caption{An example conversation comparing the multi-modal query clarification under single-turn and multi-turn scenarios.}\label{fig:fig1}
\end{figure}

Recent work~\cite{yuan2024askingmultimodalclarifyingquestions}  introduces single-turn multi-modal clarifying questions, allowing systems to present images with text. However, limiting the interactions to only one set of images limits intent inference, making it difficult to capture user needs accurately. For example, in Figure~\ref{fig:fig1}, when searching for a sofa, users need to progressively refine their preferences from general style (modern vs.\ traditional) to specific materials (leather vs.\ fabric) and finally to detailed attributes (tufted vs.\ smooth). Such natural progression in preference articulation cannot be achieved in a single turn without overwhelming users with numerous options. While existing multi-turn approaches~\cite{DBLP:journals/corr/abs-2009-11352} support dialogue flow, they lack the crucial visual context for grounding language understanding.

To address these limitations, we introduce the novel task of Multi-turn Multi-modal Clarifying Questions (MMCQ) within open-domain \ac{CS} systems. 
MMCQ enables systems to gradually refine user intent over multiple turns, where each interaction builds on the previous one by incorporating both textual questions and relevant images. This step-by-step process enhances the depth and accuracy of the clarification process, leading to more precise disambiguation of user intent and improved retrieval performance.
To facilitate research in this direction, we create a new dataset named \OurData{} that builds upon existing single-turn multi-modal clarification data~\cite{yuan2024askingmultimodalclarifyingquestions}, comprising over 13k instances of multi-turn interactions with over 14k images and 33k question-answering pairs.

Furthermore, we propose a novel ranking model, called \OurModel{} (\textbf{M}ulti-turn Multi-mod\textbf{a}l Que\textbf{r}y Clarificat\textbf{io}n), devising a two-phase ranking method to rank documents based on multi-modal conversational history. \OurModel adopts the BM25 method for initial retrieval, followed by a multi-modal generative model with a constrained generation mechanism to refine and re-rank the results. Specifically, our model leverages a pretrained multi-modal \ac{LLM} to generate the keywords sequence of relevant documents by integrating textual and visual information from the conversational interaction history.

We compare the performance of \OurModel{} against a range of models, from traditional retrieval methods to several open-sourced LLMs, and analyze the impact of multi-modal vs.\ uni-modal approaches. 
Our experiments on \OurData{} show that incorporating images in multi-turn scenarios improves MRR by up to 12.88\%  with \OurModel{}.
Additionally, a comparison between \OurData{} and its single-turn counterpart shows that multi-turn interactions consistently outperform single-turn approaches across all retrieval metrics in the multi-modal setting. Further analysis highlights \OurModel{}'s superiority, particularly in longer interactions, demonstrating the benefits of multi-turn multi-modal clarification for \ac{CS}.

\noindent
In summary, our contributions are as follows:
\begin{itemize}
    \item We introduce MMCQ as a novel task within mixed-initiative \ac{CS}, allowing the system to refine user queries through multi-turn interactions by integrating both textual and visual cues.
    \item We propose a large-scale dataset called \OurData{} to support multi-modal interactive search, which will be publicly available. We also propose \OurModel{} for effective multi-modal document retrieval in this setting. 
    \item We demonstrate the effectiveness of our model on retrieval performance by comparing it with its uni-modal and single-turn counterparts. 
\end{itemize}

\section{Related Work}
\header{Conversational question clarification.}
Query clarification improves search by refining user queries with additional context~\cite{queryclarification}, addressing ambiguities in various tasks including entity disambiguation~\cite{Coden2015DidYM}, voice-based interactions \cite{10.1145/3209978.3210160}, question answering~\cite{QAclarifyingquestion} and recommendation~\cite{recommenderCQ}.
In mixed-initiative search systems, where the conversational initiative alternates between users and agents, targeted clarifying questions have been shown to improve retrieval performance and user satisfaction~\cite{rahmani-etal-2024-clarifying,DBLP:journals/corr/abs-2410-19692}. For instance, \citet{DBLP:journals/corr/abs-2009-11352} introduced the ClariQ benchmark, which employs clarifying questions to disambiguate vague queries. Building on these foundations, \citet{yuan2024askingmultimodalclarifyingquestions} advanced the field further by developing Melon, a system that integrates visual inputs into the clarification process, thereby helping users refine their queries more effectively. Despite these advances, challenges remain in effectively merging multi-modality with multi-turn conversational interactions.

\header{Multi-modal information retrieval.} Multi-modal information retrieval has gained substantial growth by integrating different modalities to provide accurate search results \cite{MMIR}. These modalities, including text, images, audio, and video, are effective in addressing queries across diverse scenarios \cite{MMIR,yuan2024askingmultimodalclarifyingquestions}. By leveraging multi-modal data, retrieval systems can offer better and more accurate responses, which result in user satisfaction and engagement~\cite{DBLP:journals/corr/cs-IR-0311029}. Inspired by the advancements in generative large language models, new waves of multi-modal pretrained generative models have emerged which further exploit the capabilities of IR systems \cite{DBLP:journals/corr/abs-2103-00020,DBLP:journals/corr/abs-2005-09801}. Recent work has demonstrated the effectiveness of these multi-modal models in various IR tasks, such as query reformulation \cite{garg2021multimodal}, question answering \cite{Xu2019AskingCQ}, and cross-modal retrieval \cite{DBLP:journals/corr/abs-2005-09801}. Based on this, our work focuses on asking multi-modal clarifying questions in a multi-turn \ac{CS} system and investigates whether it results in better retrieval performance.

\section{Dataset Construction}
We describe how we build \OurData, our multi-turn multi-modal dataset.
\subsection{Data Collection}
Our dataset builds upon Melon~\cite{yuan2024askingmultimodalclarifyingquestions}, a single-turn dataset containing clarifying questions with images.
We use Melon's topics and facets (user information needs), which originate from TREC Web Track 2009–2012~\cite{trec2009,trec2012}, and the corresponding documents for each facet. 

\header{Multi-turn conversation synthesis.} 
We construct multi-turn conversations by systematically combining QA pairs from Melon that share the same topic. For each topic, we exhaustively generate all possible combinations of single-turn QAs to create two-, three-, and four-turn dialogues. Each turn retains its corresponding images from Melon. This approach ensures both diversity in clarification patterns and semantic coherence within each conversation.

\header{Data sampling.}
The synthesis process generates extremely large subsets for two-, three-, and four-turn conversations, with the two-turn subset alone exceeding 1 million conversations. This vast dataset poses challenges for post-processing and analysis while also containing redundant and unnatural conversations. To address this issue, we randomly sample 10,000 conversations from each subset. This selection balances dataset size while maintaining diversity and relevance.

\header{Data refinement.} 
To enhance the naturalness of synthetic data and ensure more realistic conversations, we develop an automated refinement method using GPT-4o (Algorithm~\ref{alg:refine_conversation}). While manual refinement would be ideal for ensuring conversation quality, it is impractical given our dataset scale. Our automated approach significantly reduces the required effort while maintaining high-quality dialogue refinement.

At the start of the conversation, we prompt GPT-4o to act as a user, interpreting the multi-modal conversational history and refining its responses without revealing the user's intent based on the given facet. This approach encourages a natural extension of the interaction, requiring additional exchanges to fully clarify the user’s needs. As the conversation progresses, we iteratively refine responses to gradually unveil the hidden intent, effectively simulating the natural flow of the clarification phase.
We apply this method to the filtered 30k dialogues, ensuring that the generated dialogues remain coherent and engaging while gradually revealing the hidden intent, preventing it from being disclosed too early. The detailed annotation pipeline and all prompts used are provided in Appendix~\ref{sec:appendixA}.

\begin{algorithm}[!t]
\small
\caption{Multi-turn Conversation Refinement}
\label{alg:refine_conversation}
\textbf{Input:} Conversation $d$ with $T$ turns, hidden intention $F$ \\
\textbf{Output:} Refined conversation $c$

\begin{algorithmic}[1]
\State $c \gets \{\}$ // Initialize refined conversation
\For{$t = 1$ \textbf{to} $T$}
    \If{$t == 1$}
        \State $A_t^{\text{'}} \gets \Theta_{\text{initial}}(Q_t, A_t, F)$ // $Q_t,A_t$ denote the question-answer pair at turn $t$, $\Theta$ denotes the prompting strategy
    \ElsIf{$t < T$}
        \State $A_t^{\text{'}} \gets \Theta_{\text{partial}}(Q_t, A_t, F)$ 
    \Else
        \State $A_t^{\text{'}} \gets \Theta_{\text{final}}(Q_t, A_t, F)$ 
    \EndIf
    \State $c \gets c \cup \{(Q_t, A_t^{\text{'}})\}$

\EndFor
\end{algorithmic}
\end{algorithm}

\subsection{Quality Control}

To validate the quality of our synthetic dataset, we conducted a human evaluation assessing four key metrics: \textit{relevance}, \textit{coherence}, \textit{diversity}, and \textit{intent reveal}. These metrics were chosen to evaluate critical aspects of our dataset construction process, where single-turn QA pairs from the Melon dataset \cite{yuan2024askingmultimodalclarifyingquestions} were combined and refined into multi-turn dialogues.
Given our dataset's scale, we randomly sampled 10\% of the topics for annotation. Two of the authors independently evaluated 178 conversations using a 5-point Likert scale (1: poor to 5: excellent) (detailed definition of the metrics see Appendix \ref{sec:appendix-quality}).
Our human evaluation results (Table~\ref{tab:data_quality}) demonstrate the effectiveness of our construction approach. Relevance scores show consistent improvement from Turn 1 (3.62, $\sigma$=1.29) to Turn 4 (4.11, $\sigma$=0.97), validating our GPT-4o refinement strategy for maintaining topical focus. While coherence (3.36, $\sigma$=1.10) indicates some minor inconsistencies, the strong diversity score (4.01, $\sigma$=0.97) confirms that our sampling strategy captured varied aspects of topics without repetition. Most notably, the high intent completion score (4.65, $\sigma$=0.87) validates our approach of gradually revealing user intent across turns.
These results prove that our data generation pipeline successfully produces well-structured and semantically rich multi-turn conversations, making ClariMM a valuable resource for training multi-turn multi-modal retrieval systems.

\begin{table}[!t]
    \centering
    \small
       \setlength{\tabcolsep}{4pt}
    \begin{tabular}{lccc}
        \toprule
        \textbf{Metric} & \textbf{Mean} & \textbf{Std Dev} & \textbf{Median} \\
        \midrule
        Relevance (Turn 1) & 3.62 & 1.29 & 3.00 \\
        Relevance (Turn 2) & 3.56 & 1.24 & 3.00 \\
        Relevance (Turn 3) & 3.78 & 1.09 & 3.00 \\
        Relevance (Turn 4) & 4.11 & 0.97 & 4.00 \\
        Coherence          & 3.36 & 1.10 & 3.00 \\
        Diversity          & 4.01 & 0.97 & 4.00 \\
        Intent reveal & 4.65 & 0.87 & 5.00 \\
        \bottomrule
    \end{tabular}
    \caption{Human evaluation scores for relevance, coherence, diversity, and intent reveal.}
    \label{tab:data_quality}
\end{table}

\subsection{Dataset Statistics}

\begin{table}[!t]
\centering
\small
\begin{tabular}{lr}
\toprule
\textbf{Metric} & \textbf{Value} \\ \midrule
\# topics & 298 \\ 
\# facets & 1,070 \\ 
\# all questions & 4,969 \\ 
\# images & 14,869 \\ 
\# answers & 33,477 \\
\# 2-Turn Conversations & 7,782 (59.36\%) \\ 
\# 3-Turn Conversations & 3,391 (25.86\%)\\ 
\# 4-Turn Conversations & 1,935 (14.78\%) \\ 
\bottomrule
\end{tabular}
\caption{Statistics of the \OurData{} dataset.}
\label{tab:dataset_stats1}
\vspace{-6mm}
\end{table}
\begin{figure*}[t]
    \centering
    \includegraphics[width=\linewidth]{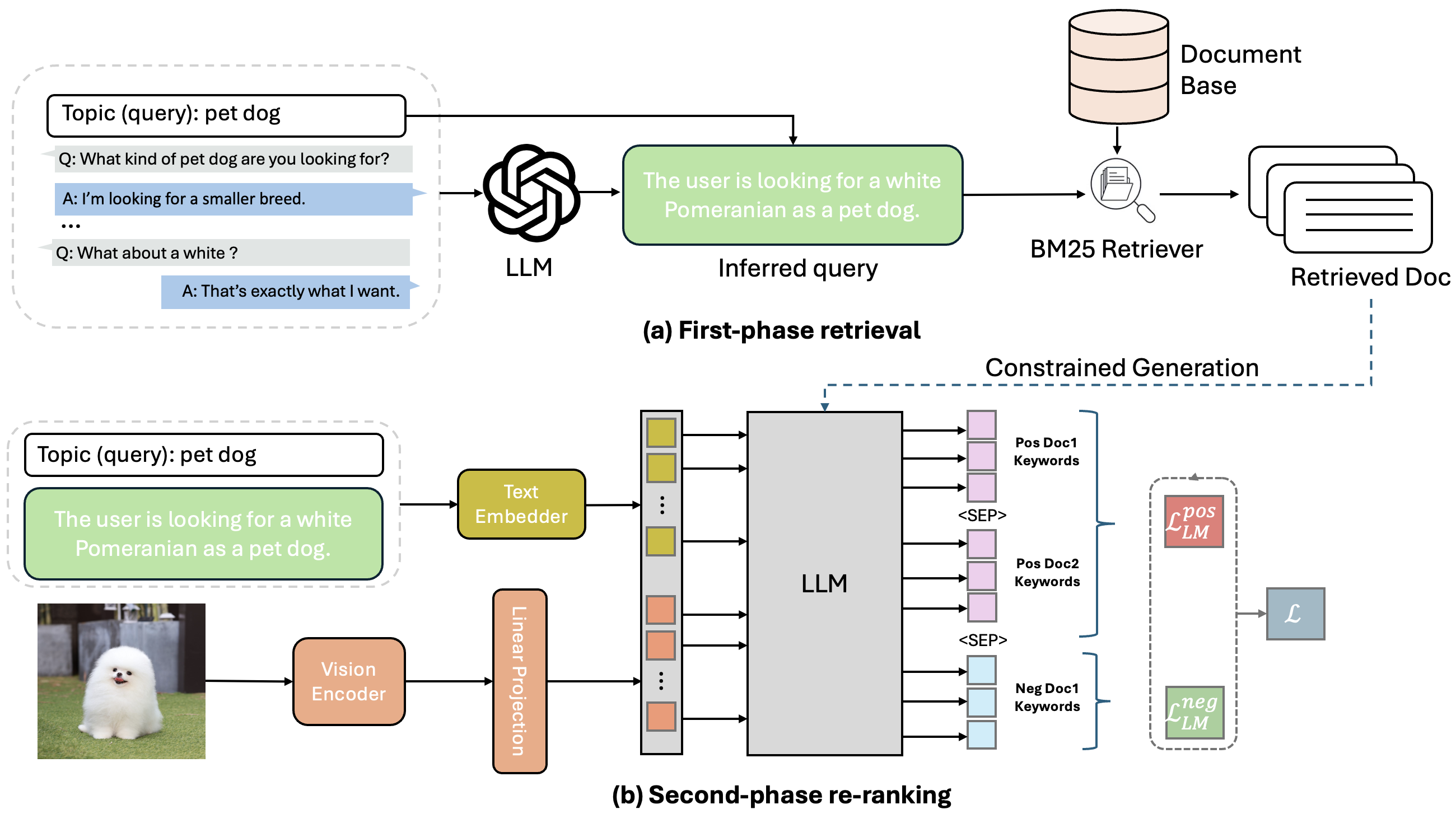} 
    \caption{Overview of the \OurModel{} two-phase retrieval framework.}
    \label{fig:framework}
    \vspace{-4mm}
\end{figure*} 
Table~\ref{tab:dataset_stats1} provides an overview of the basic statistics of \OurData{}. The dataset comprises a total of 298 search topics and 1070 facets. It consists of 4,969 clarifying questions accompanied by 14,869 images, resulting in an average of 2.99 images per question. 
Additionally, the dataset includes 33,477 answers and every question has its own answer. Overall, the dataset consists of over 7k two-turn conversations, 3k three-turn conversations, and 1k four-turn conversations.

\section{Our Method} 
\subsection{Problem Formulation}

Following \cite{yuan2024askingmultimodalclarifyingquestions}, we consider a set of topics denoted as \( T = \{t_1, t_2, \dots, t_k\} \), serve as user queries. Each topic \( t_i \) is associated with a set of facets, defined as \( F_i = \{ f_i^1, f_i^2, \dots, f_i^{n_i} \} \), where \( n_i \) represents the number of facets for \( t_i \). Each facet \( f_i^j \) captures a distinct aspect of \( t_i \), specifying a particular user information need.  
Given a topic \( t \) and an information need (facet) \( f \), the user engages in a conversation \( C \) consisting of \( k \) turns. Each conversation comprises a sequence of \textbf{multi-modal} clarifying questions \( Q = \{q_1, q_2, \dots, q_k\} \) and their corresponding \textbf{text-only} answers \( A = \{a_1, a_2, \dots, a_k\} \). Each question \( q_i \) consists of text and may optionally include some images. At the end of each conversation, a set of documents $D$ is retrieved and ranked based on the conversation. The goal is to determine the hidden facet $f$ and learn a retrieval function $R(\cdot)$ that maps the conversation context and topic to a ranked list of documents, such that \( R(C,t) \to D \).

\subsection{Framework Overview}
As shown in Figure~\ref{fig:framework}, we propose a framework called \OurModel{} to retrieve relevant documents given the multi-modal conversational history (details see Section \ref{sec:gen_retrieval}). The process begins with the system receiving the user's query as input. It then refines the query by incorporating the conversation history to generate an inferred query. Next, BM25 is applied for first-phase retrieval, retrieving the top 100 most relevant documents. Then, we introduce a multi-modal generative re-ranking model that incorporates the inferred query to refine and re-rank the initial results. Specifically, we train the model to generate keywords for the top relevant documents, leveraging both textual and visual information. By incorporating multi-modal information, the model effectively re-ranks the retrieved documents to enhance relevance.

\subsection{Multi-modal Two-phase Retrieval}
\label{sec:gen_retrieval}
\subsubsection{First-phase Retrieval}
In the first phase of our retrieval process, we employ BM25 to retrieve an initial set of relevant documents from the document base given the query $t$ and conversational history context $C$. Since \( C \) is lengthy and might contain noise, we extract an inferred query \(\Phi\) from \( C \) using GPT-4o (prompts see Appendix \ref{sec:appendiC}). Given the inferred query \( \Phi\), the set of retrieved documents is obtained as:
\begin{equation}
    D_{initial} = \text{BM25}(t, \Phi, D)~,
\end{equation}
where $D$ is the initial document set and $D_{initial}$ is the first-ranked result. The retrieved documents are then passed to subsequent stages for further refinement and re-ranking using multi-modal information with generative models.

\subsubsection{Multi-modal Re-ranking}
To integrate multi-modal information, we propose a generative re-ranking model based on a multi-modal LLM.

\noindent \textbf{Image and text encoding.}
Our model encodes the input image \( I \) using the SigLIP \cite{siglip} vision encoder $f_{\text{img}}$ to extract image feature $\mathbf{z}$: $\mathbf{z} = f_{img}(I)$. The image feature is then projected into the LLM's embedding space using a learned projection matrix \( W \) and concatenated with the text embedding \(\tau\), where \(\tau\) is obtained from the text embedder $f_{text}$: $\tau=f_{text}(t,\Phi)$. The final output $\mathbf{e}$ is then computed as:
\begin{equation}
    \mathbf{e} = f_{\text{LLM}}([W \mathbf{z}; \tau ])~,
\end{equation}
where \( f_{\text{LLM}} \) is the LLM responsible for generating the final re-ranking output.

\noindent \textbf{Keyword extraction.} 
Following previous work in generative retrieval~\cite{rankedGR,li2023learningrankgenerativeretrieval}, we train the multi-modal LLM to generate a ranked sequence of document IDs. Each document $d$ is identified by a unique keyword-based ID denoted as \( K_d \), ensuring efficient retrieval and semantic relevance. Specifically, we extract five representative keywords per document using YAKE~\cite{YAKE}. These keywords serve as compact semantic descriptors that capture each document's core information.

\noindent \textbf{Model training.}
We train the model to generate a ranked sequence of document IDs based on the multi-modal input \( x \), refining the initial BM25 ranking \( D_{\text{initial}} \). To improve the model's ability to distinguish between good and bad ranking results, we train it to generate keywords for relevant and irrelevant documents sequentially, with individual documents separated by a [SEP] token. Relevant and irrelevant samples are identified based on their overlap with the ground-truth labels in \( D_{\text{initial}} \). For the loss function, we use the Margin Ranking Loss for ranking which is defined as:
\begin{equation}
    \mathcal{L}_{\text{rank}} = \max(0, m + \mathcal{L}_{\text{LM}}^{\text{pos}} - \mathcal{L}_{\text{LM}}^{\text{neg}})~,
\end{equation}
where $m$ is the margin, $\mathcal{L}_{\text{LM}}^{\text{pos}}$ and $\mathcal{L}_{\text{LM}}^{\text{neg}}$ are the language modeling loss for the relevant and irrelevant samples respectively. In detail, the language modeling loss can be represented as:
\begin{equation}
\mathcal{L}_{\text{LM}} = - \sum_{t=1}^{T} \log P_\theta(y_t \mid y_{<t}, x)
\end{equation}
where \(y_{<t}\) denotes the sequence of tokens generated before time step \(t\), and \(\theta\) represents the model parameters.

The final loss is a combination of the positive sample's language modeling loss and the ranking loss: 
\begin{equation}
    \mathcal{L} = \mathcal{L}_{\text{LM}}^{\text{pos}} + \lambda_{\text{rank}} \cdot \mathcal{L}_{\text{rank}}~,
\end{equation}
here $\lambda_{\text{rank}}$ is the weighting factor.

\header{Inference.}
During inference, to prevent the model from generating arbitrary tokens, we employ a constrained generation technique 
~\cite{constraineddecoding} to ensure that only valid keywords are selected and generated. That is, we restrict the model vocabulary to a predefined set of allowed keywords from $D_{initial}$.
Specifically, at each decoding step \( t \), let the current partial sequence be \( y_{<t} \). We define the allowed set of tokens $A_t$ as:
\begin{equation}
\{ v \in \mathcal{V} \mid \exists z \in \mathcal{T}, \text{s.t. } y_{<t} \oplus v =\text{prefix}(z) \}~,
\end{equation}
where \(\mathcal{V}\) is the vocabulary, \(\mathcal{T}\) is the trie encoding for all valid keyword sequences, and \(\oplus\) denotes sequence concatenation. By masking the probability distribution for the next token to consider only those in \( A_t \), we ensure that the generated output adheres strictly to the allowed keywords.

\section{Experiments}
\begin{table*}[!h]

    \centering
    \setlength{\tabcolsep}{3mm}
    
    \begin{adjustbox}{max width=1\textwidth}
    \begin{tabular}{l cccccccc}
    \toprule

         & Img. & MRR & P@1 & P@3 & P@5 & nDCG@1 & nDCG@3 & nDCG@5  \\
    \midrule
    BM25 & \xmark & 50.74 & 39.62 & 36.16 & 36.03 & 25.80 & 23.39 & 24.56  \\
    Bert & \xmark & 56.36 & 46.08 & 41.50 &  41.37 & 35.70 & 33.82 & 34.01 \\
    T5 & \xmark & 52.15 & 41.30 &  37.64 & 38.63 &  41.30 &  38.82 & 39.39 \\
    Qwen-2 & \xmark &  46.48 & 42.26 & 39.72 & 39.23 & 40.08 & 37.96 & 36.88\\
    \midrule
    VisualBert & \cmark & 56.50 &  46.57 & 46.24 & 44.02 &  35.33 & 36.65 &  36.28 \\
    VLT5 & \cmark  & 53.22 & 42.34 & 38.83 & 39.43 & 42.34 & 39.90 & 40.26 \\
    \OurModel & \cmark & \textbf{59.36} & \textbf{48.10} & \textbf{47.09} & \textbf{45.48} & \textbf{46.90} & \textbf{45.77} & \textbf{43.98}   \\
 \bottomrule
    \end{tabular}
    \end{adjustbox}
\caption{Experimental results (\%) on multi-turn conversations.}
    \label{tab:mainexperiment}
\vspace{-2mm}
\end{table*}

\begin{table*}[!h]

    \centering
    \setlength{\tabcolsep}{3mm}
    \begin{adjustbox}{max width=1\textwidth}
    \begin{tabular}{l cccccccc}
    \toprule

         & Img. & MRR & P@1 & P@3 & P@5   & nDCG@1 & nDCG@3 & nDCG@5  \\
    \midrule
    BM25 & \xmark & 42.94 & 32.07 & 30.81 & 30.37 & 20.39 & 20.15 & 21.02 \\ 
    Bert & \xmark & 49.34 & 39.22 & 37.42 & 36.27 & 29.66 & 29.42 & 29.13 \\ %
    T5 & \xmark & 41.37 & 28.08 & 28.97 & 28.88 & 28.08 & 29.16 & 31.92 \\
    Qwen-2 & \xmark & 44.30 & 40.56 & 37.68 & 35.97 & 38.40 & 35.94 & 33.68 \\
    \midrule
    VisualBert & \cmark & 45.95 & 37.75 & 33.50 & 32.55 & 28.43 &25.83 & 25.20 \\ 
    VLT5 & \cmark  & 43.18 & 30.46 & 28.92 & 28.94 & 30.46 & 29.69 & 30.42 \\
    \OurModel & \cmark & \textbf{53.24} & \textbf{46.54} & \textbf{43.48} & \textbf{40.02} & \textbf{41.85} & \textbf{39.56} & \textbf{38.68} \\
 \bottomrule
    \end{tabular}
    \end{adjustbox}

\caption{Experimental results (\%) on single-turn conversations.}
\label{tab:mainexperiment-singleT}
\vspace{-4mm}
\end{table*}

\subsection{Experimental Setup}
We split \OurData{}'s facets into 80\% for training, 10\% for validation, and 10\% for testing, and create the corresponding datasets accordingly. As a result, the training set consists of 9,688 conversations and 856 facets, while the validation and testing set each contains 672 conversations and 107 facets. To create the single-turn comparison set, we adopt only the first turn of each conversation and we obtain the inferred query as input. We choose LLaVA-OneVision-7b as the base model for \OurModel{}. For retrieval evaluation, we employ Mean Reciprocal Rank (MRR), Precision (P@k), and Normalized Discounted Cumulative Gain (nDCG@k) where $k \in \{1,3,5\}$. The ground truth relevance documents are sourced from the TREC Web Track 2009-2012 \cite{trec2009,trec2012}. All hyperparameters are detailed in Appendix~\ref{sec:appendixB}. We report the performance of Oracle image selection. Our experiments are conducted using the PyTorch framework, with training and evaluation performed on one NVIDIA V100 and two NVIDIA A100 GPUs.

\subsection{Compared Methods}

We first adopt several uni-modal baselines by removing image information from the model input to simulate a text-only interaction.

\begin{description}

    \item [BM25]~\cite{Robertson2009BM25} ranks documents based solely on the text input, without any re-ranking.
    \item [Bert] \cite{Devlin2019BERTPO} reranks the BM25 results with Bert model. We adopt the implementation from \citet{MacAvaney2019CEDRCE}.
    \item [T5] \cite{DBLP:T5} is trained to perform re-ranking by generating keyword sequences of relevant documents given a query. We use the \verb|T5-base| version in our experiment.

    \item [Qwen-2]~\cite{qwen2} 
 ranks documents similar to T5, but uses \verb|Qwen-2-7b| as the base model.
    \end{description}

We also compare our method with several multi-modal baselines:
\begin{description}

    \item [VisualBert] \cite{Li2019VisualBERT} is a multi-modal model with Bert structure and is trained with pairwise softmax loss for re-ranking.
    
    \item [VLT5] \cite{Cho2021VLT5} takes multi-modal input and is trained to output the keyword of the documents with constrained generation.
   
\end{description}

\subsection{Experimental Results}
We report the performance of multiple baselines on multi-turn and single-turn settings in Table \ref{tab:mainexperiment} and \ref{tab:mainexperiment-singleT}. We observe that language-model-based rankers such as T5 and Bert outperform the traditional lexical method BM25. 
We further analyze the impact of incorporating images in the document retrieval task. Our findings indicate that using images enhances retrieval performance, particularly in multi-turn conversations, compared to models that rely solely on text. For instance, in the multi-turn case, VLT5 achieves a P@1 of 42.34\%, outperforming its uni-modal counterpart T5, which records a P@1 of 41.30\%. These results highlight the advantage of multi-modal information in capturing a more comprehensive user intent over longer conversational histories. However, this benefit diminishes in the single-turn scenario where we see a 1.47\% decrease in P@1 comparing Bert with VisualBert. This is due to the image being misleading in the first turn, as the model benefits less from visual information when there is limited context.
Results further show that all models perform notably better in multi-turn conversations than in single-turn ones, as added context helps capture user intent more effectively.
Notably, \OurModel{} consistently outperforms the other baselines in the multi-turn and single-turn settings, achieving the highest scores across key metrics and emphasizing its superior ability to leverage contextual cues.

\section{Extensive analysis}
\subsection{Impact on different turns}
\begin{figure}[t]
  \includegraphics[width=\columnwidth]{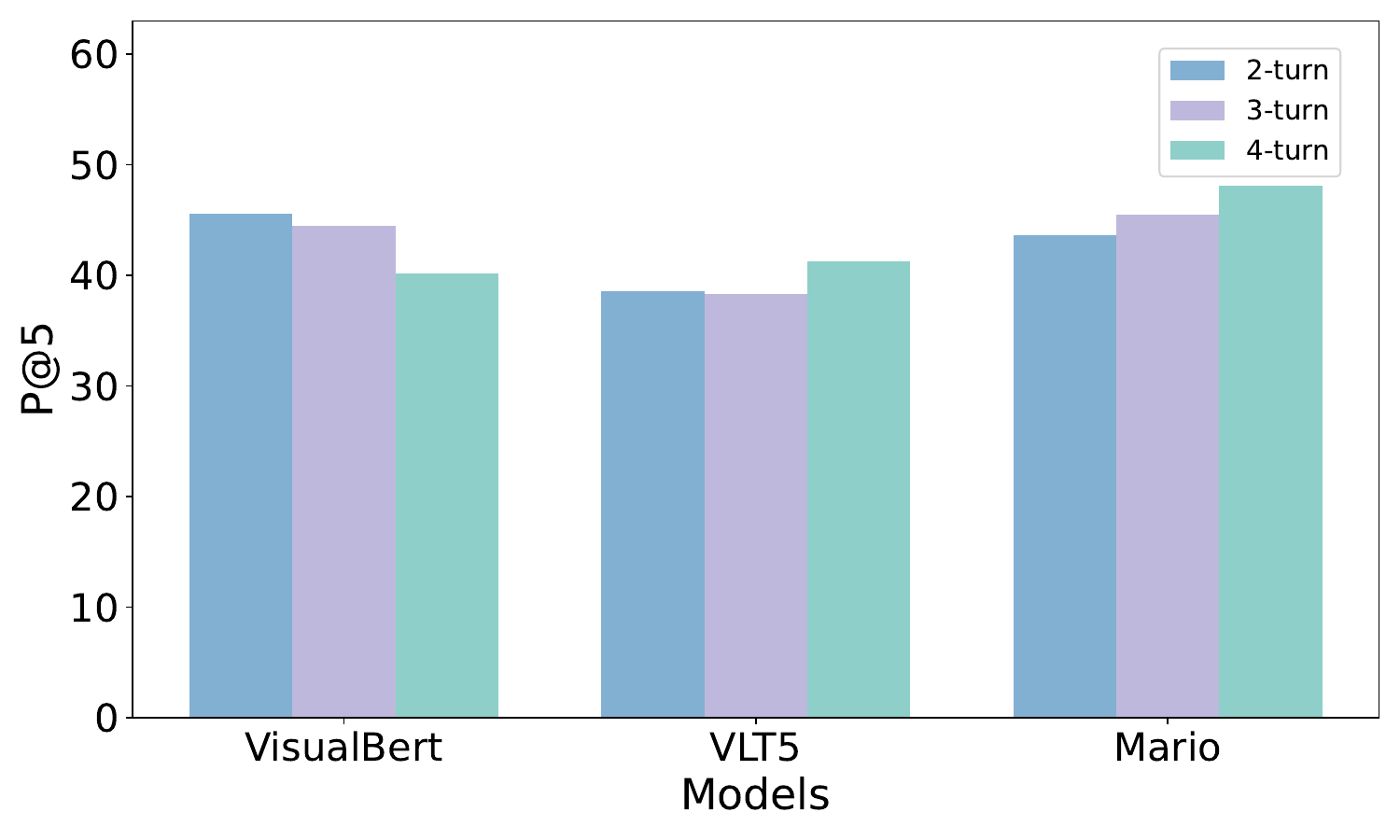}
  \caption{P@5 scores under different turn counts in \OurData{}.}
  \label{fig:fig3}
  \vspace{-4mm}
\end{figure}

We further report the retrieval performance under the different number of turns for VLT5, VisualBert, and \OurModel{} in Figure~\ref{fig:fig3}. As shown in the figure, VLT5 indicates only a modest improvement from 38.59 (two-turn) to 41.30 (four-turn), indicating limited gains from the additional turns. VisualBert's performance even declines as the conversation length increases, starting at 45.58 for two-turn data and dropping to 40.19 for four-turn data. This suggests that VisualBert struggles to leverage the increasing context effectively in longer conversations.
In contrast, \OurModel{} demonstrates consistent and substantial improvements with each additional turn, with P@5 increasing from 43.60 (two-turn) to 48.12 (four-turn). This significant gain confirms that \OurModel{} excels in multi-turn conversational retrieval and outperforms VLT5 and VisualBert in longer interactions. This highlights the model’s ability to effectively capture the evolving intent and incorporate context across turns making it particularly well-suited for handling long conversations.

\newcolumntype{C}[1]{>{\centering\arraybackslash}m{#1}}

\begin{table*}[!h]
\centering
\renewcommand{\arraystretch}{1.2}
\setlength{\tabcolsep}{3mm}
\begin{adjustbox}{max width=\textwidth}
{
\begin{tabular}{
    C{0.8cm}    %
    C{2.5cm}    %
    C{4cm}|%
    C{1.8cm}    %
    C{4cm}      %
    C{2.5cm}    %
    C{3cm}    %
}
    \toprule
    \textbf{Idx} & \textbf{Topic} & \textbf{Facet} & \textbf{Turn Num} & \textbf{Inferred Query} & \textbf{Image} & \textbf{Image Effect} \\
    \midrule

    \multirow[c]{6}{*}{1}  
      & \multirow[c]{6}{*}{Teddy bears}
      & \multirow[c]{6}{*}{%
         \parbox{4.5cm}{\raggedright
           Find giant teddy bears
         }
      }
      & \begin{minipage}[c][2cm][c]{1.8cm}
          \centering
          multi-turn
        \end{minipage}
      & \begin{minipage}[c][2cm][c]{4cm}
          \raggedright
          Looking for giant teddy bears
        \end{minipage}
      & \begin{minipage}[c][2cm][c]{2.5cm}
          \centering
          \includegraphics[width=1.8cm]{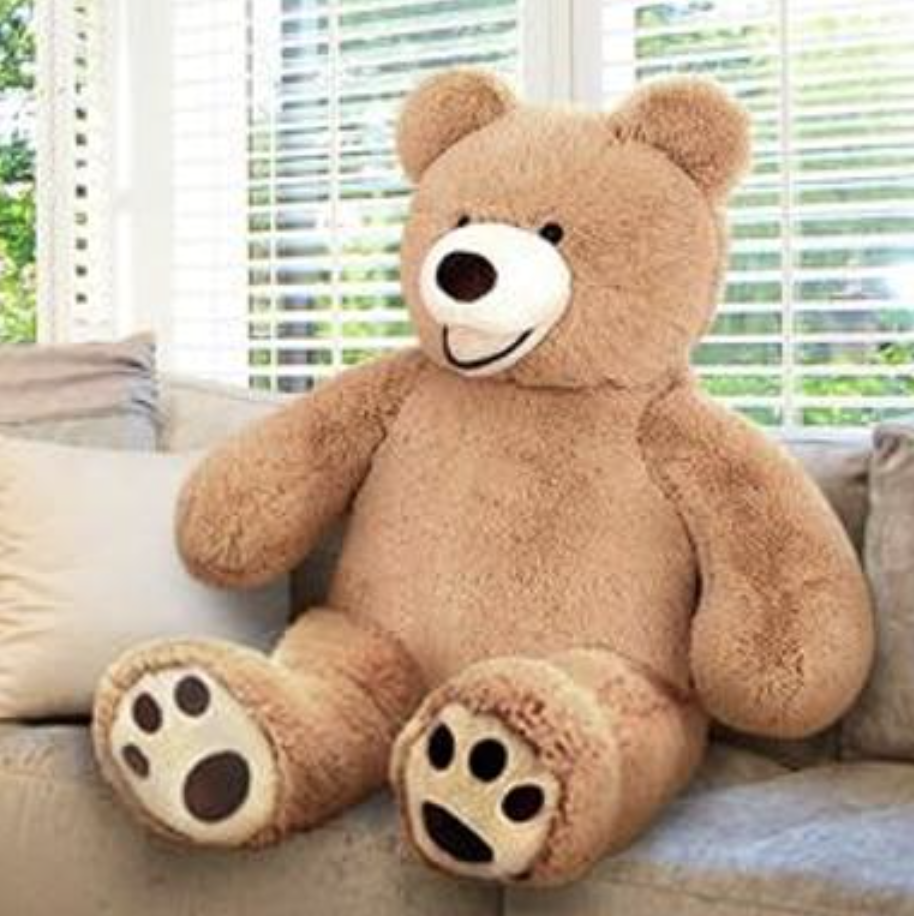}
          \vspace{0.1cm}
        \end{minipage}
      & +0.2
    \\ \cline{4-7}
      & & &
      \begin{minipage}[c][2cm][c]{1.8cm}
        \centering
        single-turn
      \end{minipage}
      & \begin{minipage}[c][2cm][c]{4cm}
        \raggedright
        Exploring options related to teddy bears
      \end{minipage}
      & \begin{minipage}[c][2cm][c]{2.5cm}
        \centering
        \vspace{0.1cm}
        \includegraphics[width=1.8cm]{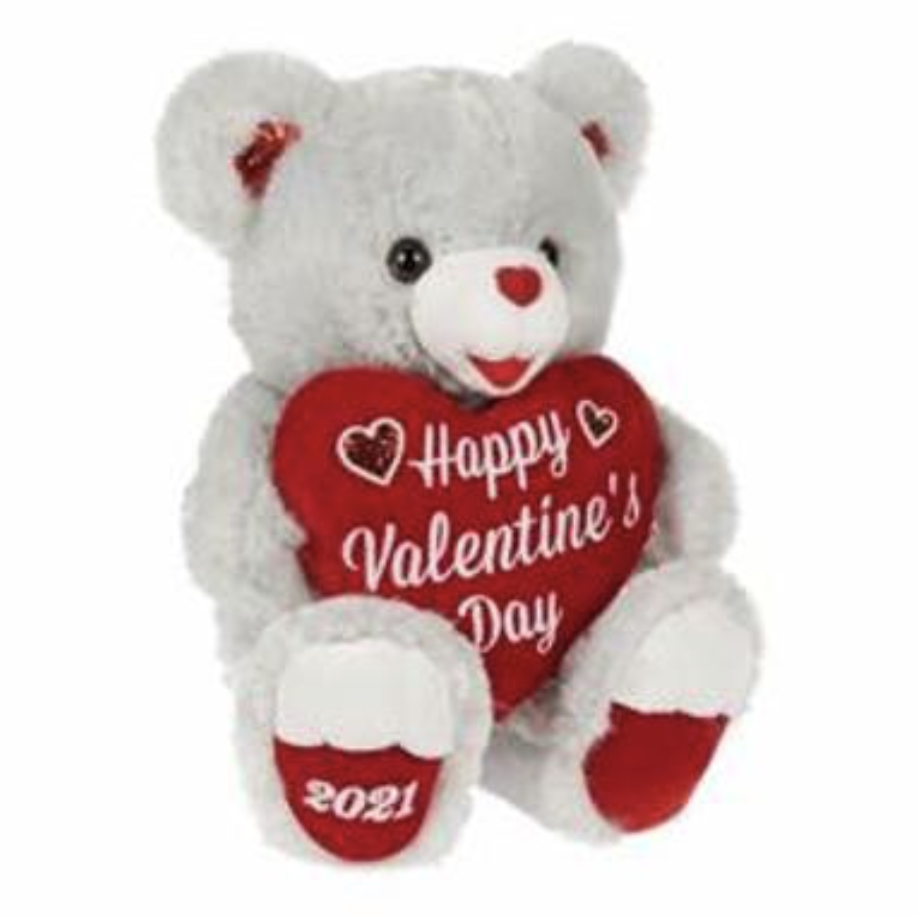}
      \end{minipage}
      & 0
    \\
    \midrule

    \multirow[c]{6}{*}{2} 
      & \multirow[c]{6}{*}{Hobby Stores} 
      & \multirow[c]{6}{*}{%
         \parbox{4.5cm}{\raggedright
           Where can I buy radio-controlled planes?
         }
      }
      & \begin{minipage}[c][2cm][c]{1.8cm}
          \centering
          multi-turn
        \end{minipage}
      & \begin{minipage}[c][2cm][c]{4cm}
          \raggedright
          Places to buy radio-controlled planes
        \end{minipage}
      & \begin{minipage}[c][2cm][c]{2.5cm}
        \centering
        \includegraphics[width=1.8cm]{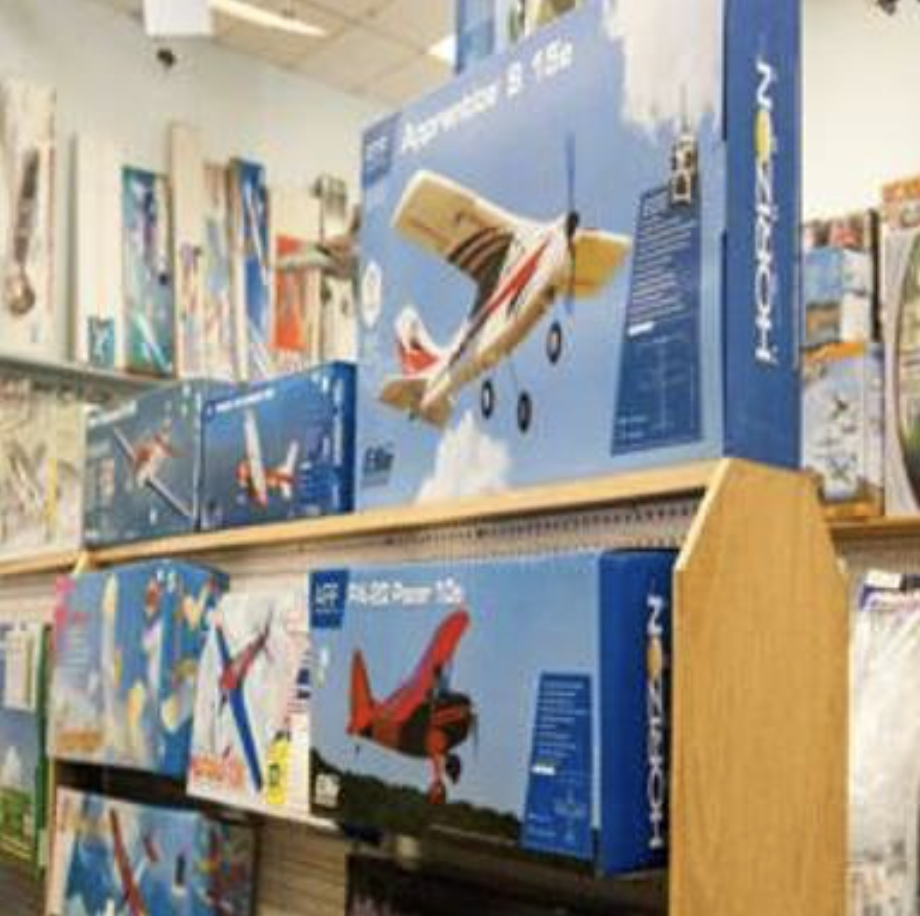}
        \vspace{0.1cm}
      \end{minipage}
      & +0.8
    \\ \cline{4-7}
      & & &
      \begin{minipage}[c][2cm][c]{1.8cm}
        \centering
        single-turn
      \end{minipage}
      & \begin{minipage}[c][2cm][c]{4cm}
        \raggedright
        Finding a new hobby
      \end{minipage}
      & \begin{minipage}[c][2cm][c]{2.5cm}
        \centering
        \vspace{0.1cm}
        \includegraphics[width=1.8cm]{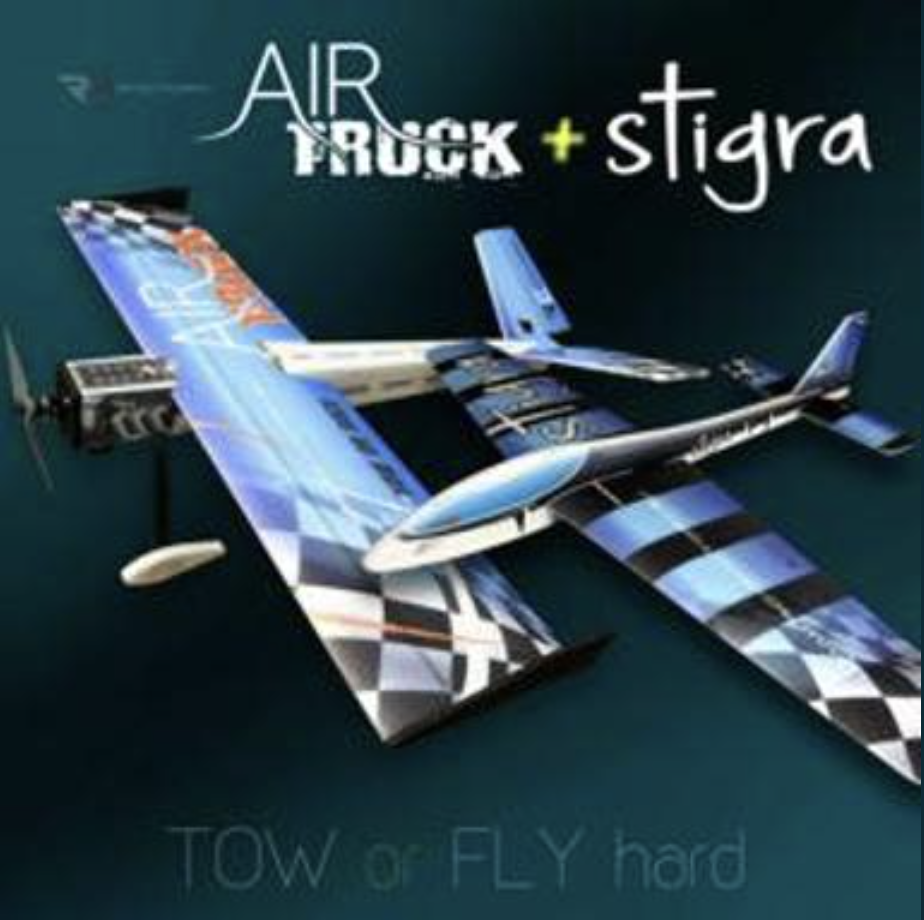}
        
      \end{minipage}
      & +0.2
    \\
    \midrule

    \multirow[c]{6}{*}{3} 
      & \multirow[c]{6}{*}{Wilson's Disease} 
      & \multirow[c]{6}{*}{%
         \parbox{4.5cm}{\raggedright
           What are the symptoms of Wilson's disease?
         }
      }
      & \begin{minipage}[c][2cm][c]{1.8cm}
          \centering
          multi-turn
        \end{minipage}
      & \begin{minipage}[c][2cm][c]{4cm}
        \raggedright
        Understanding symptoms of Wilson's disease
      \end{minipage}
      & \begin{minipage}[c][2cm][c]{2.5cm}
        \centering
        
        \includegraphics[width=1.8cm]{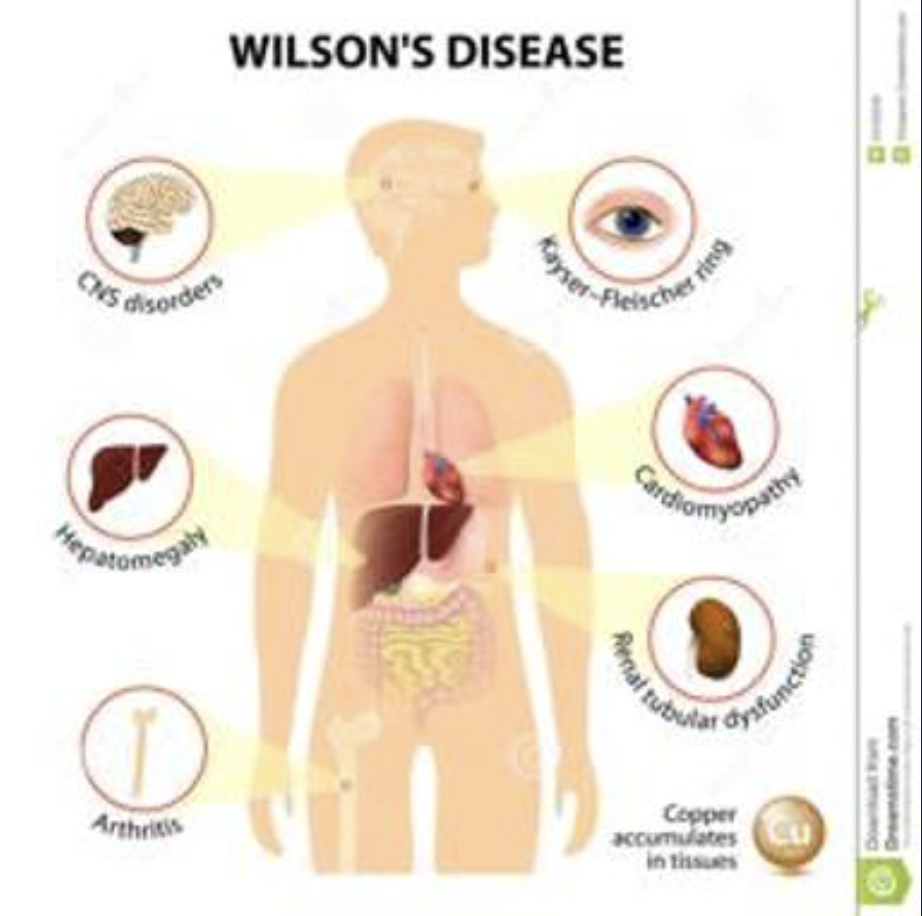}
        \vspace{0.1cm}
      \end{minipage}
      & +0.2
    \\ \cline{4-7}
      & & &
      \begin{minipage}[c][2cm][c]{1.8cm}
        \centering
        single-turn
      \end{minipage}
      & \begin{minipage}[c][2cm][c]{4cm}
        \raggedright
        Understanding the condition of Wilson's disease
      \end{minipage}
      & \begin{minipage}[c][2cm][c]{2.5cm}
        \centering
        \vspace{0.1cm}
        \includegraphics[width=1.8cm]{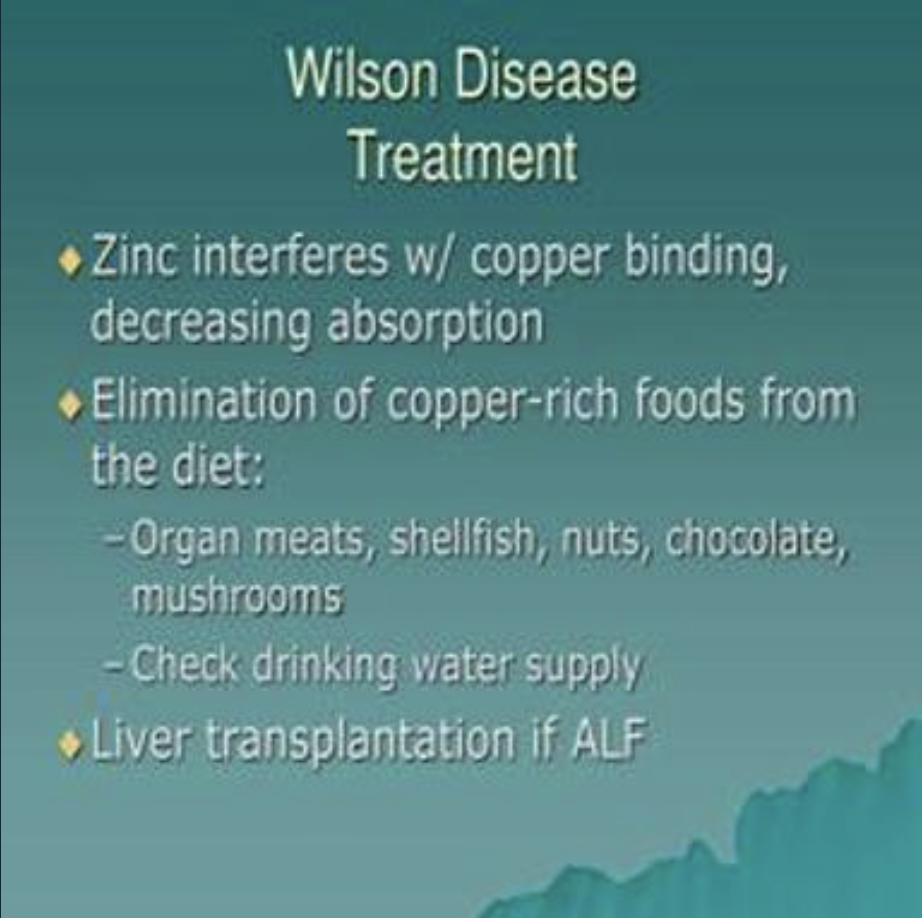}
      \end{minipage}
      & -0.4
    \\
    \bottomrule
\end{tabular}
}
\end{adjustbox}
\caption{Case study on \OurModel{}. A positive Image Effect indicates an increase in performance after adding the image, while a negative effect indicates a performance drop.}
\label{tab:casestudy}
\vspace{-4mm}
\end{table*}

\begin{table}[!ht]
\centering
\begin{tabular}{lccccc}
\hline
\multicolumn{1}{c}{\multirow{2}{*}{\textbf{Method}}}     & \multicolumn{2}{c}{\textbf{Seen}} & & \multicolumn{2}{c}{\textbf{Unseen}} \\
\cline{2-3} \cline{5-6}
& MRR & P@5 & & MRR & P@5 \\
\hline
Bert & 54.55 & 40.31 & & 51.50 & 34.00\\
T5 & 53.12 & 34.23 && 38.55 & 24.16 \\
VisualBert & 53.53 & 39.46 & & 51.85 & 35.17 \\
VLT5  & 55.31 & 34.46 & & 43.35 & 25.49 \\
\OurModel{} & \textbf{58.68} & \textbf{46.17} &  & \textbf{57.79} & \textbf{43.23}\\
\hline

\end{tabular}
\caption{Comparison between seen and unseen topics.}
\label{tab:seen-unseen}
\vspace{-4mm}
\end{table}

\subsection{Impact on different topics}

We further evaluate the performance of various models on seen and unseen topics to evaluate their robustness and generalization capabilities.  We re-split the \OurData{} dataset into training, unseen, and seen testing sets. The unseen testing set consists of 10\% of all topics, entirely excluded from the training process. In contrast, the seen testing set includes topics that are also present in the training set.
As shown in Table~\ref{tab:seen-unseen}, Bert-based models (\textit{i.e.}, VisualBert \& Bert) and our model demonstrate a relatively consistent performance across both seen and unseen topics, with minimal differences in evaluation metrics. T5-based models (\textit{i.e.}, VLT5 \& T5), however, show a more significant decline between the seen and unseen sets, which suggests greater sensitivity to new topic distributions. Furthermore, we observe that the impact of using images in the unseen topics is more noticeable than in the seen topics. We can see a 4.8\% increase in MRR when comparing T5 and VLT5 on unseen data, however, this difference is smaller (2.29\%) on the seen domain. This suggests that incorporating visual information provides a greater advantage when dealing with unfamiliar topics.

\subsection{Case study}
To demonstrate the effect of adding images to the multi-turn and single-turn conversations, we perform a case study in Table \ref{tab:casestudy}. In most cases, including images provides valuable contextual information which enhances the model's performance. Notably, adding images in multi-turn conversations tends to have a more significant positive effect compared to single-turn cases. For example, in case 2, adding an image in the multi-turn setting improves the P@5 score by 0.8 whereas adding an image in the single-turn scenario only boosts P@5 by 0.2. However, there are instances where images can negatively impact performance. In case 3, the inferred query from the single-turn conversation focuses on understanding the condition of Wilson's disease. Unfortunately, due to the insufficiency of the inferred query, the returned image fails to align with the user's hidden intent, as it includes treatment-related information. The user is primarily interested in learning about the symptoms of this disease, not its treatment and this image leads to a negative impact on the P@5 score. By contrast, in the multi-turn scenario, the image displays symptoms, thereby providing valuable information that enhances the model's performance.

\section{Conclusion}
We investigate the novel task of asking multi-modal clarifying questions in multi-turn \ac{CS} systems. To enable research in this domain, we introduce a large-scale dataset named \OurData{}, which contains over 13k multi-turn multi-modal interactions and 33k question-answer pairs, accompanied by 14k images. We also propose a multi-modal query clarification framework named \OurModel{}, which adopts a two-phase retrieval strategy by combining initial BM25 ranking with a multi-modal generative re-ranking model. We further compare \OurModel{} with state-of-the-art models. Experimental results show that multi-turn multi-modal interactions significantly help users refine their queries, leading to improved retrieval performance.

\section*{Limitations}
Several limitations remain for future work. First, we synthetically developed our dataset from Melon, which despite our best efforts to refine it for realism, may not fully capture the spontaneity and complexity of true user interactions. Future work could address this limitation by leveraging techniques like data augmentation or reinforcement learning from human feedback (RLHF) to bridge the gap between synthetic and natural interactions. Additionally, it remains an open question how much images truly enhance the user experience in the MMCQ task. Since the effectiveness of visual information can depend heavily on its contextual relevance and the specific user intent, our current approach might not optimally handle ambiguous or noisy visual inputs. Future work should explore methods to better integrate and disambiguate visual data to maximize their contribution to the overall user experience.

\section*{Ethical Statement}
All images and user topics in our dataset are sourced from publicly available datasets, ensuring that no private or sensitive information is included. The collection and use of these resources strictly comply with the terms of use and licensing agreements set by the original dataset providers. 
We have diligently verified that all materials originate from public sources, conducting our research with the highest regard for data privacy and ethical integrity.

\bibliography{acl_latex,anthology}

\appendix

\section{Dataset Creation and Prompts}
\label{sec:appendixA}

\begin{figure*}[!t]
    \centering
    \includegraphics[width=\linewidth]{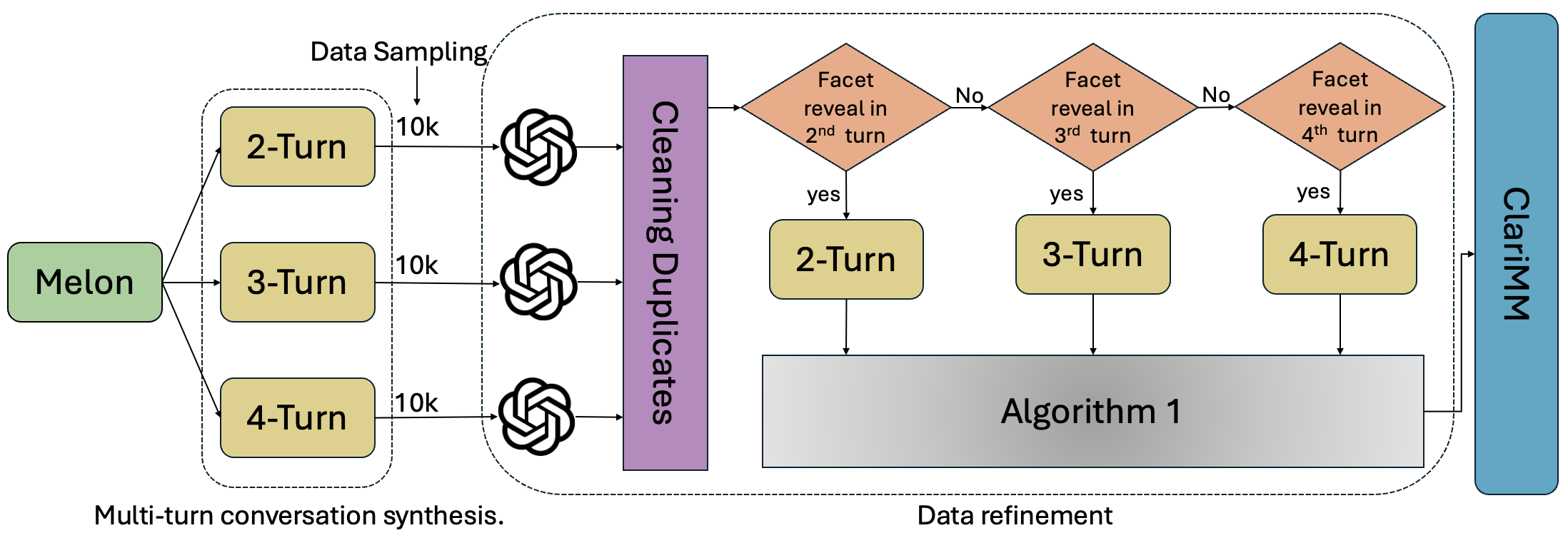}
    \caption{Dataset creation pipeline.}
    \label{fig:dataset_creation}
    \vspace{-1.5mm}
\end{figure*}

We use a multi-step refinement process, as shown in Figure \ref{fig:dataset_creation} to address the unnaturalness of synthetic data. We first prompt GPT-4o to determine if two QA convey similar information in a single conversation, then we remove entries identified as having duplicate QA structures using Prompt A in Table \ref{tab:All_Prompts}. This step helps detect and remove redundant or highly similar QAs.

Next, We prompt GPT-4o to analyze each conversation turn and identify whether the hidden facet intention is revealed prematurely using prompt B in Table \ref{tab:All_Prompts}. This Prompt assesses whether the hidden facet intention is revealed too early. It judges whether a provided answer can be interpreted as the same as the facet intention. For instance, If the conversation has four turns and the hidden intention is revealed in the second turn, we extract those two turns and add them to the two-turn dataset.

As illustrated in the figure, the four-turn data undergoes the most rigorous filtering process compared to the two-turn and three-turn data, which explains its lower count in Table \ref{tab:dataset_stats1}. Consequently, the amount of available data decreases as the number of turns increases because, in most cases, the intention is revealed prematurely.  

Finally, we introduce an additional refinement step using Algorithm \ref{alg:refine_conversation} to ensure the conversational flow is as realistic as possible. In this algorithm, we employ three prompts, $\Theta_{\text{initial}}$, $\Theta_{\text{partial}}$, and $\Theta_{\text{final}}$, using 2-shot learning. In Table \ref{tab:All_Prompts} we show that these prompts iteratively refine responses to gradually unveil the hidden intent to effectively simulate the natural progression of the clarification phase.

\section{Quality Control Metric}
\label{sec:appendix-quality}
The following metrics were used to assess the quality of \OurData{} during human evaluation: \textbf{Relevance}: Each turn's alignment with the original topic (assessed per turn); \textbf{Coherence}: Logical flow between combined QA pairs (assessed per dialogue); \textbf{Diversity}: Variation in responses and avoidance of redundancy (assessed per dialogue); and \textbf{Intent reveal}: Effectiveness of progressive intent revelation (assessed per dialogue).

\section{Hyperparameter Settings}
\label{sec:appendixB}
Our code is based on PyTorch \cite{pytorch} and Huggingface Transformers \cite{HuggingFace}. For Llava-OneVision, we use the 7b pretrained version, 1e-4 as the learning rate and 2 for the batch size. For the loss function, we set the margin to 2.0 and \(\lambda_{rank}\) to 0.75.For generation, we set the number of beams to 10. For first-phase document retrieval, we retrieved the top 100 documents using BM25.

\section{Inferred Query Extraction}
\label{sec:appendiC}
To capture the user’s intent from a multi-turn conversation, we employ a summarization step using GPT-4o that focuses on what the user is actually interested in. It compresses the conversation into a short “inferred query” discarding irrelevant details such as off-topic remarks. By isolating only the essential user request, the system can more effectively guide subsequent retrieval ensuring that the user’s primary goal remains at the forefront.

\begin{table}[!h]
    \centering
    \setlength{\tabcolsep}{3mm}
    \begin{adjustbox}{max width=1\columnwidth}
    {\LARGE
    \begin{tabular}{l}
    \toprule
    Prompt \\
    \midrule
    Extract the user's intent based on the conversation. \\ Only mention what they are interested in. \\
     Conversation: \{conversation\} \\
    
    \bottomrule
    \end{tabular}
    }
    \end{adjustbox}
\caption{Prompts used for dataset creation.}
\label{tab:query_extract}
\end{table}

\begin{table*}[!h]
    \centering
    \setlength{\tabcolsep}{3mm}
    \begin{adjustbox}{max width=1\textwidth}
    {\LARGE
    \begin{tabular}{l| l}
    \toprule

       Type  & Prompt Content \\
    \midrule
    Prompt A &  I will provide you with two pairs of questions and answers. Determine if these two question-answer pairs \\&contain similar information. Output "yes" or "no" and explain why. \\
    &Question 1: \{question1\} Answer 1: \{answer1\}, Question 2: \{question2\} Answer 2: \{answer2\} \\
    \midrule

    Prompt B & I will provide you a pair of question-answer and a facet (user's hidden intention).\\& Judge whether the answer aligns with the facet intention. If yes, generate: "intention reached". \\
    & Facet intention: \textit{facet\_intention}, Question: \textit{question}, Answer: \textit{answer} \\
    \midrule

    Prompt $\Theta_{\text{initial}}$ & \textbf{Examples:}\\
    & Example 1: \\
    & \textbf{Facet:} How to fix a car engine.\\& \textbf{Question:} Do you want to buy a car? \textbf{Answer:} No, I am not looking to buy a car.\\[5pt]
    & Example 2: \\
    & \textbf{Facet:} Find coffee shops near me. \\&\textbf{Question:} Would you like to make a cup of coffee? \textbf{Answer:} No, thank you, I want to buy one.\\[5pt]
    & I provided you with some examples above. Now, modify the following answer so that it is connected to the question\\&and doesn't reveal the hidden intention of the facet like in the examples. Ensure your answer doesn't violate the facet.\\[5pt]
    & \textbf{Prompt:}\\
    & Imagine you are a user answering an agent question. Modify this answer without revealing any hidden intention\\& of the facet and without violating the facet.\\[5pt]
    & Facet: \{facet\}, Question 1: \{question1\}, Answer 1: \{answer1\}, \{examples\}   \\
    \midrule
    
    Prompt $\Theta_{\text{partial}}$ & \textbf{Examples:}\\
    & Example 1: \\
    & \textbf{Facet:} The user wants to buy a red car.\\&\textbf{Question:} Are you looking for a specific color? \textbf{Answer:} I am considering a color, but I haven't decided fully yet.\\[5pt]
    & Example 2: \\
    & \textbf{Facet:} I'm looking for the car-part.com website.\\&\textbf{Question:} Do you want to sell used car parts? \textbf{Answer:} For now, I am mainly focused on finding a website.\\[5pt]
    & I provided you with some examples above. Now, modify the following answer to reveal only a partial abstract of the \\& hidden intention (facet) and hint at the user's interests without revealing the full intention\\[5pt]
    & \textbf{Prompt:}\\
    &Imagine you are a user answering an agent question. Modify the following answer to reveal \\&only a partial abstract of the hidden intention (facet). Do \textbf{NOT} reveal the full hidden intention.\\[5pt]
    & Facet: \{facet\} Question 3: \{question2\} Answer 3: \{answer2\} \{examples\} \\
    \midrule

    Prompt $\Theta_{\text{final}}$ & \textbf{Examples:}\\
    & Example 1: \\
    & \textbf{Facet:} The user wants to buy a red car.\\&\textbf{Question:} Are you looking for a specific color? \textbf{Answer:} Yes, I am looking for a red car to buy.\\[5pt]
    & Example 2: \\
    & \textbf{Facet:} I'm looking for the car-part.com website.\\&\textbf{Question:} Do you want to sell used car parts? \textbf{Answer:} No, I am just looking for the car-part.com website.\\[5pt]
    & I provided you with some examples above. Now, modify the following answer to fully reveal the hidden intention \\ & in a clear and direct manner, and ensure that the answer reflects the facet without ambiguity.\\[5pt]
    & \textbf{Prompt:}\\
    &Imagine you are a user answering an agent question. Modify the following answer to fully reveal the hidden facet.\\& Ensure that the answer clearly reflects the facet.\\[5pt]
    & Facet: \{facet\}, Question 3: \{question3\}, Answer 3: \{answer3\}, \{examples\}  \\
    \bottomrule
    \end{tabular}
    }
    \end{adjustbox}
\caption{Prompts used for dataset creation.}
\label{tab:All_Prompts}
\end{table*}

\end{document}